\documentclass[rnote]{aa}

\usepackage{graphics}
\usepackage{epsfig}
\usepackage{latexsym}
\begin{document}
\hyphenation{brems-strah-lung}
\title{\emph{INTEGRAL} and \emph{Swift}/XRT observations of  IGR~J19405-3016}

\author{ Shu Zhang\inst{1}, Yu-Peng Chen\inst{1}, Diego F. Torres\inst{2}, Jian-Min Wang\inst{1,3}, Ti-Pei Li\inst{1,4}, Jun-Qiang Ge\inst{1}
}

\institute{Key Laboratory for Particle Astrophysics, Institute of High
Energy Physics, Beijing 100049, China
\and
ICREA \& Institut de Ci\`encies de l'Espai (IEEC-CSIC), Campus UAB, Facultat de Ci\`encies, Torre C5-parell, 2a planta, 08193 Barcelona, Spain 
\and
Theoretical Physics Center for Science Facilities (TPCSF), CAS
             \and
Center for Astrophysics,Tsinghua University, Beijing 100084, China
          }

\offprints{Shu Zhang}
\mail{szhang@mail.ihep.ac.cn}

\date{Received  / Accepted }

\titlerunning{\emph{INTEGRAL} and \emph{Swift}/XRT observations ...}
\authorrunning{Shu Zhang et al.}

  \abstract
  {}
{IGR J19405-3016 is reported in the 3rd IBIS catalog as  one of its lowest  significance sources ($\sim$ 4.6 $\sigma$ under an exposure of about 371 ks).
This leads to a caveat in multi-wavelength studies, although the source was identified in the optical as a  Seyfert 1.  The currently available \emph{INTEGRAL} data on the source have increased to an exposure time of $\sim$ 1400 ks, which stimulates us to investigate the reality of this source again  by using all the available data from \emph{INTEGRAL} and \emph{Swift}/XRT. }
{We analyzed all available observations  carried out by the International Gamma-Ray Astrophysics
Laboratory (\emph{INTEGRAL})  on  IGR J19405-3016.
 The data were processed by using the latest version OSA 7.0. 
In addition, we  analyzed all the available \emph{Swift}/XRT data on this source.}
{We find that IGR J19405-3016 has a detection significance of $\sim$ 9.4 $\sigma$ in the 20-60 keV band during the observational period between March 2003 and March 2008.
This confirms a real source detection reported previously.  The source position and error location are therefore updated. The source is found to be constant over years at the hard X-rays. 
We analyzed the \emph{Swift}/XRT observations on IGR J19405-3016 as well, and    find  that the  spectrum can be  fitted with a simple power law model.  Over the three XRT observations, the source flux varied by  up to 39\% from the average, and the spectrum is generally soft.  The combined XRT/ISGRI spectrum is well fitted with a simple power law model (photon index  2.11$\pm$0.03) with a column density fixed at 8.73$\times$10$^{20}$ atoms/cm$^2$. Such a photon index is  consistent with the mean value 1.98 (dispersion 0.27)  obtained from \emph{Swift}/BAT AGN samples at 14-195 keV.  The spectral slope of IGR J19405-3016 is softer than the average spectral slope found  elsewhere.
A similar discrepancy is found with other results regarding Seyfert 1 AGNs. A possible explanation for this simple spectral description may be that the low level of the column density allows for the `true' spectrum to appear at soft X-rays as well.
} 
{} 

   \keywords{ X-rays: individual: IGR J19405-3016}

   \maketitle

\section{Introduction}

Four hundred twenty-one sources are listed in the 3rd IBIS/ISGRI catalog 
(Bird et al. 2007). Among them,  171 are Galactic accreting systems, 
122 extragalactic objects, and 113 are sources of unknown nature. Most of the unclassified sources were identified through optical and near-infrared spectroscopy  (Masetti et al. 2004, 2006a, 2006b, 2006c, 2006d, 2008, 2009; Chaty et al. 2008; Nespoli et al. 2008). For the newly identified sources, about 55\% are AGNs (almost equally divided into Seyfert 1s and 2s), 32\% are X-ray binaries, and 12\% are CVs (Masetti et al. 2009).  The optical identifications are performed by first searching in the error location of \emph{INTEGRAL}/ISGRI  for a possible counterpart at the soft X-ray observations and, accordingly, the error location can be largely improved to a level that allows for looking into optical data for the final identification. IGR~J19405-3016 is one of the \emph{INTEGRAL} weakly identified sources studied by Masetti et al. (2008) in their most recent optical identification campaign.

IGR~J19405-3016 was first reported as a low-significance source in the 3rd IBIS/ISGRI catalog (Bird et al. 2007). The source was detected at 4.6 $\sigma$ level under an exposure of $\sim$ 371 ks in the energy band 20-40 keV. The lowest significance for a source in Bird's catalog is 4.5 $\sigma$. Therefore, IGR~J19405-3016 is among the few sources that were reported by \emph{INTEGRAL}/ISGRI with the lowest significances. Bird et al. (2007) point out that the  catalog sources
detected with significance above 5 $\sigma$ have a probability of less than 1 percent of being spurious, and  detections with lower significance have a higher probability of being unreal. Accordingly, although  a Seyfert 1.2 was found as an optical counterpart  around IGR~J19405-3016 by Masetti et al. (2008), a caveat was claimed therein  in the sense that the hard X-ray detection itself may be spurious, and hence the proposed optical identification may not correspond to an actual counterpart  (Masetti et al. 2008).

After the discovery of the source by \emph{INTEGRAL}, \emph{Swift}/XRT  detected a source at soft X-rays within the IBIS error location (Landi et al. 2007a). The XRT observation locates the source at RA(J2000) = 19h 40m 15.15s and Dec(J2000) = -30d 15m 48.5s,  with a 3.5 arcsecond uncertainty  at 90\% confidence
level. The X-ray spectrum is  modeled with an unabsorbed power law with a photon index $\sim$ 2.1  (Landi et al. 2007a).  Given  that an additional exposure will  lead to a large increment in source significance and that the currently available \emph{INTEGRAL} observations around IGR~J19405-3016 have been accumulated  to roughly 1400 ks, it is appropriate to re-investigate   the hard X-ray properties of   IGR~J19405-3016. Also, after the report of Landi et al. (2007a), further \emph{Swift}/XRT observations  (roughly 1.6 ks more) were available in 2008, which allow for further investigation on  the source  at  soft X-rays.
In this paper we report the results of our \emph{INTEGRAL} and \emph{Swift}/XRT analyses on IGR J19405-3016.

\section{Observations and data analysis}

 \emph{INTEGRAL} (Winkler et al. 2003) is a 15 keV - 10 MeV $\gamma$-ray mission. The main instruments  
are the Imager on Board the INTEGRAL Satellite (IBIS, 15 keV - 10 MeV; Ubertini et al. 2003) and the SPectrometer onboard INTEGRAL (SPI, 20 keV - 8 MeV; Vedrenne et al. 2003).
 They are supplemented by
the Joint European X-ray Monitor (JEM-X, 3-35 keV) (Lund et al. 2003) and the Optical
Monitor Camera (OMC, V, 500-600 nm) (Mas-Hesse et al. 2003).  At the lower energies (15 keV - 1 MeV), 
the CdTe array ISGRI (Lebrun et al. 2003) of IBIS has a better continuum sensitivity than SPI. 
The satellite was launched in October 2002
into an elliptical orbit with a period of 3 days. Due to the coded-mask
design of the detectors, the satellite normally operates in
dithering mode, which suppresses the systematic effects on spatial
and temporal backgrounds.

The \emph{INTEGRAL} observations were carried out in the 
so-called individual  SCience Windows (SCWs), with a typical time duration of about 2000 seconds each.
 Only IBIS/ISGRI public data were  taken
into account, because the source is too weak to be detected by JEMX and SPI. 
 The available \emph{INTEGRAL} observations, when IGR J19405-3016 fell into
 the 50$\%$ coded field of view of ISGRI (offset angle less than 10
degrees),  comprised about 566
SCWs, adding up to a total exposure time of $\sim$ 1400 ks (until March 22, 2008). 
IGR J19405-3016 therefore had roughly  1000 ks of exposure more than  used in the previous report (Bird et al. 2007). 
 The details of the analyzed \emph{INTEGRAL} observations on IGR J19405-3016, 
including the exposure and the time periods, 
are summarized in Table 1. 
Most of these observations were carried out in the 5x5 dithering mode. We subdivided the data into 4 groups according to the 
observational sequence. The other observational modes of the selected data are staring mode and Hexagonal mode, the sum of which comprise only 5.6\% of the whole data. Data from these two modes can just as well be used to produce a mosaic map without introducing any systematic errors. This has been confirmed through our consulting the INTEGRAL help desk.  
The data reduction was  performed by using the standard Online Science Analysis (OSA)
software version 7.0, the latest released version. The results 
were obtained by running the pipeline from the flowchart to the image level and the spectrum level. 
The flux and the detection significance were derived in the mosaic map at the source position  revised in this work.

 \emph{Swift} (Gehrels et al. 2004) is a $\gamma$-ray burst explorer  launched November 20, 2004. 
It carries three co-aligned detectors:  the Burst Alert Telescope 
(BAT, Barthelmy et al. 2005), the X-Ray Telescope (XRT, Burrows et al. 2005),  
and the Ultraviolet/Optical Telescope (UVOT, Roming et al. 2005). 
 We took only
\emph{Swift}/XRT data into account, because BAT data were not available.
 The XRT uses a grazing incidence Wolter I telescope to focus X-rays onto 
a state-of-the-art CCD. XRT has an effective area of 110 cm$^2$,  an  FOV of 23.6 arcminutes, 
an angular resolution (half-power diameter) of 15 arcseconds, and it operates in the 0.2-10 keV energy range,  providing 
the possibility of extending the investigation on the  source to soft X-rays.

There are three \emph{Swift} snapshots available for IGR J19405-3016,  each one  with an exposure over  1 ks.  
 The observations were carried out in the photon-counting mode, with exposures of 7.6 ks (ID 00036657004, on July 31, 2007), 5.9 ks (ID 00036657005, on August 7, 2007) and   1.6 ks (ID 00036657005, on June 7, 2008), respectively. The first two observations were reported in Landi et al. (2007a).  See details in Table 2.
 We analyzed the \emph{Swift}/XRT 0.3-7 keV data by using the latest released analysis software, provided in HEAsoft 
version 6.4.   The XRT data reduction follows those described in Landi et al. (2007b). Here the source events were extracted within a circular region of radius $\sim$ 40 pixels, centered on the source position. A radius of 20 pixels (corresponds to 47 arcseconds) encloses about 90 percent of the PSF at 1.5 keV (Capalbi et al. 2005). A larger radius   extracts more source counts, hence improves the statistics in spectral fittings. The
spectra were fitted with XSPEC v12.3.1  (Dorman \& Arnaud 2001) and the model parameters 
estimated  at 90$\%$ confidence level. 

\section{Results}
\subsection{\emph{INTEGRAL} }

\begin{figure}[ptbptbptb]
\centering
\includegraphics[angle=0, scale=0.4]{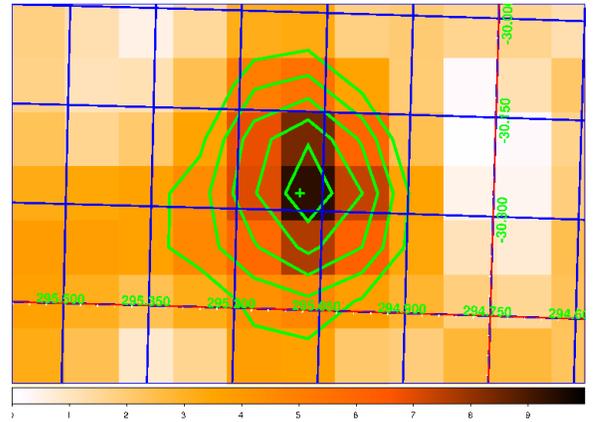}
      \caption{The 20-60 keV significance map of IGR J19405-3016 as obtained 
from the combined  revolutions 0056 - 0664 (MJD 52729-54547).   The contours start at a 
significance level of 4 $\sigma$ with steps of 1 $\sigma$.}
         \label{ima_isgri}
\end{figure}

 The imaging analyses show that the best source detection, $\sim$ 9.4 $\sigma$ at 20-60 keV (Fig. \ref{ima_isgri}), was derived in the mosaic map of  all data. 
The source is not detectable at lower energies by JEMX or at higher energies by SPI.
Such a detection is much more significant than the previous report of a 4.5 $\sigma$ signal with  an exposure of 371 ks,  at 20-40 keV (Bird et al. 2007). The source flux is about 1.40$\pm$0.15 
 mCrab in the 20-60 keV band over the period 2003-2008.  Under such a high-significance detection at 20-60 keV, the source position is improved to RA/Dec (J2000) = 295.0895$^{\circ}$/-30.2732$^{\circ}$, with a radius  of 3.4  arcminutes in error circle ( 90$\%$ confidence level), also improved from the previous report of 5.4 arcminutes (Bird et al. 2007). 

We also investigated  the source detection in individual observational groups as listed in Table 1, in the 20-60 keV band. The source is detected in Revs. 0056-0244 (MJD=52729-53291) with a significance of 4.9 $\sigma$ and a flux of 1.5$\pm$0.3 mCrab; in Revs. 0258-0371 (MJD=53332-53672) with a significance of 3.5 $\sigma$ and a flux of 1.4$\pm$0.4 mCrab and Revs. 0416-0498 (MJD=53805-54050) with a significance of 7.8 $\sigma$ and a flux of 1.5$\pm$0.2 mCrab. One sees that the source is detectable in each observational group and the flux remains constant at about 1.5 mCrab level over years at hard X-rays.  The ISGRI data after Revs. 0498 consist of only 16 scws (Revs.543-664), which are about 17 ks  exposure in total. The source was not detected with this exposure, and  we have a 2-$\sigma$ flux upper limit of 3.3 mCrab at 20-60 keV.  
If, during Revs. 543-664, the source kept
at the same flux level as that of the previous revolutions, it would have only 
been 'clearly detected' at 1-$\sigma$ level. See details for the results in Table 1. 
The source is  detected in the 4 adjacent energy bands of 20-25,  25-30,  30-40, and 40-60 keV, with detection significances of 4.0, 5.4, 6.1, and 4.2 $\sigma$, respectively,  by combining all the ISGRI data. If there is a constant source at hard X-rays, an exposure of 1400 ks can improve the detection significance by a factor of 1.95 with respect to   the previous report (Bird et al. 2007), which means  a detection significance of $\sim$ 9 $\sigma$, similar to  what we obtain from the summed data. In short, all  results derived by us indicate a real and steady source.

\subsection{\emph{Swift}/XRT}

 The \emph{Swift}/XRT imaging analysis  shows  a single source  detected at soft X-rays within the error location of IGR J19405-3016.   Figure 2 is the  \emph{Swift}/XRT image produced by 
 combining the three XRT observations to show the most accurate location of the source at soft X-rays. The position is obtained as RA(J2000) = 19h 40m 15.00s and Dec(J2000) = -30d 15m 48.6s, with an error radius of   3.5 arcseconds  at 90\% confidence level, which is  consistent with the report in Landi et al. (2007a). 
Over plotted are the positions of IGR J19405-3016, as derived in Bird et al. (2007) and  in the present work, and their error circles. It is obvious that the revised position and improved error circle are more indicative of the correlation between the soft X-ray source  detected with XRT and the hard X-ray source IGR J19504-3016. 

To look into the properties of the XRT source at soft X-rays, we  carried out the spectral analysis of the three XRT observations. We find that data from each of the three observations can be fitted by using a simple power law model. The data in ID 00036657004 and ID 00036657005 need an additional component of absorption, with a column density derived as 5-6 $\times$ 10$^{20}$ atoms/cm$^2$, but this is not necessary for the data from observation ID 00036657006, probably  due to a low exposure of only $\sim$ 1.6 ks. 
 The source was clearly detected in XRT observation of 2008 June as well. The relatively low flux level and low exposure in this observation led to larger errors in model parameters.
 The reduced $\chi^2$ were derived with values of $\le$ 1.0. The column density as measured  directly from the XRT spectra of IGR J19405-3016  is rather low: a value around 6$\times$10$^{20}$atoms/cm$^2$ can be even comparable to the  Galactic 
column density obtained with the web version  (9.15$\times$10$^{20}$atoms/cm$^2$)\footnote{\texttt{http://heasarc.gsfc.nasa.gov/cgi-bin/Tools/W3nh/w3nh.pl}} and from Dickey $\&$ Lockman (1990) (8.73$\times$10$^{20}$atoms/cm$^2$), measured $\sim$ 0.4 deg away from IGR J19405-3016. We therefore fix the column density at 8.73$\times$10$^{20}$atoms/cm$^2$ and fit again the XRT data with a power law model. We  have the fittings with the reduced  $\chi^2$ derived around 1.1 (see Table 2).   We find that  the  XRT fluxes  in these snapshots can vary up to 39\%  with respect to the average at soft X-rays. The overall spectrum is relatively  soft irrespective of the change in flux. The summed XRT data can be well-fitted by using a simple power law model, with a  reduced  $\chi^2$ $\sim$ 1.13  for 198 dof\footnote{degrees of freefom}. This suggests that the spectral evolution, if any, should be not very strong over the three XRT snapshots. 
See Table 2 for details of the spectral results.

\subsection{Combined XRT/ISGRI spectrum}
The  summed XRT spectrum is combined with the ISGRI spectrum extracted  from the all \emph{INTEGRAL} observations.   The joint spectrum can be fitted with a  model of simple power law plus fixed absorption (reduced $\chi^2$$\sim$1.14  for 202 dof) (Fig. 3).  The photon index $\Gamma$ and  normalization were derived as   2.11$\pm$0.03   and 4.70$\pm$0.10$\times$10$^{-3}$  ph cm$^{-2}$ s$^{-1}$ keV$^{-1}$, respectively.   A constant is introduced to account for the difference in normalization between \emph{Swift}/XRT and \emph{INTEGRAL}/ISGRI and the sporadic snapshots of XRT observations, and is derived as 2.19$^{+0.65}_{-0.60}$, which is slightly more than unity. Such deviation can be ascribed to flux variability at soft X-rays observed in the sporadic snapshots of XRT. 

\section{Discussion and summary}

As listed in the 
3rd IBIS catalog (Bird et al. 2007), IGR J19405-3016 belongs to a group of   sources with significance less than 5 $\sigma$.  
 About 10-20 percent of them might result from false detections (Bird et al. 2007). 
By taking the currently available \emph{INTEGRAL} observations, the exposure of which is roughly a factor of four more than  used in  the 3rd IBIS catalog (Bird et al. 2007),  we find IGR J19405-3016  at a significance level of $\sim$ 9.4 $\sigma$ at 20-60 keV 
in the sum of the observations between March 2003 and March 2008. The source is consistently detected at significance level $\ga$ 4 $\sigma$, in the three observational groups covering a time period of almost 5 years and in the adjacent four energy bands between 20-60 keV. We therefore confirm the  detection of a real source in a previous report (Bird et al. 2007). Hence the caveat mentioned in Masetti et al. (2008) in optical identification of IGR J19405-3016 has disappeared.  A similar result is shown  as well in Beckmann et al. (2009), where a significance of 11.8 $\sigma$ was reported  for IGR J19405-3016 in 18-60 keV.  Given the distance of a redshift z=0.052 (Masetti et al. 2008), the source luminosity is estimated as 2.5$\times$10$^{44}$ ergs/s (1-100 keV, use $H_0$=70 km/s/Mpc, $q_0$=0, $\Omega_\lambda$=0.73).  Our analysis shows that the source was rather stable over 5 years at energies above 20 keV. The source location and error circle are updated accordingly, which are more indicative of a source  observed by \emph{Swift}/XRT at soft X-rays as its real counterpart.

 At soft X-rays we find  that IGR J19405-3016 has a relatively soft spectrum and the XRT flux can vary up to  39\%  with respect to the average at soft X-rays.  Given  the low sensitivity of ISGRI (about 4-5 mCrab for individual SCW) and low average flux level at $>$ 20 keV ($\sim$1.5 mCrab for IGR J19405-3016),  such a variability at hard X-rays cannot be studied because of the lack of sufficient signal-to-noise  ratio in the ISGRI data, although  short variability cannot be excluded.  In Table 1 one sees that the source is stable on a long time scale at hard X-rays.  It has been reported several times that, for Seyfert 1, their fluxes  can be less variable at hard X-rays than at soft X-rays (Beckmann et al. 2007; Molina et al. 2009; Gliozzi et al. 2003).   A possible explanation for this is that the large number of the scatterings of the soft X-rays in the corona region can let the variability be washed out at hard X-rays. Such an idea was proposed in  Gliozzi et al. (2003), but so far  the detailed modelings are not yet  available.

 The combined XRT/ISGRI energy spectrum is well-fitted using a model of simple power law plus fixed absorption component. The spectral index  falls into the range of the spectral index of the AGN measured by \emph{Swift}/BAT at 14-195 keV band (Tueller et al. 2008). The analysis of the first 9 months of the data of \emph{Swift}/BAT survey of AGN resulted in 103 AGN being detected in 14-195 keV.  The average  spectral index of these samples is 1.98, with an  rms of 0.27.  For IGR J19405-3016, the spectral index of 2.11$\pm$0.03  is consistent with the media value from BAT.  About 74 BAT AGNs have the archival spectrum at soft X-rays. A comparison of these samples between hard and soft X-rays shows that the BAT spectral slope is in general 0.23 steeper than in the soft X-rays (Tueller et al. 2008).  A possible explanation for this could be that the spectrum is  closer to the intrinsic one at hard X-rays than at soft X-rays due to the influence of the   material local to AGN (Nandra et al. 1999; Tueller et al. 2008). The broad-band (1-100 keV) analysis of 36  Seyfert 1 AGNs, detected by INTEGRAL in the 20-40 keV band, presents a power law shape  with an average  photon index of 1.7  and a dispersion of 0.2 (Molina et al. 2009).  The photon index of 2.11$\pm$0.03 derived for IGR J19405-3016 in $\sim$ 1-100 keV is   thus slightly more than the result of Molina et al. (2009).  A similar discrepancy is found with the results of Beckmann et al. (2009) regarding Seyfert 1 AGNs. The column density as measured in IGR J19405-3016 is  rather low, and from the X-ray spectral data, there is no need to introduce a further absorption local to the AGN.  
A possible explanation for the derived spectral  properties of IGR J19405-3016  may be that the low level of the column density allows for the `true'  spectrum to also be detected  at soft X-rays, and thus
the overall measured spectral shape of the source is actually the
true one.

 In summary, the most recent analysis carried out with the latest software releases confirms that the previous  weak detection,  hence  the low-significance source IGR J19405-3016 as listed in Bird et al. (2007), is not spurious. As a result, the optical identification of a Seyfert 1.2 by Masetti et al.  (2008) is strengthened. A spectral index $\sim$ 2.11 supplemented with a low absorption may  indicate that the `true' spectrum of IGR J19405-3016  extends back to soft X-rays.

\begin{table}[ptbptbptb]
\begin{center}
\label{tab1}
\caption{ IBIS/ISGRI observations log of IGR J19405-3016. }
\vspace{5pt}
\small
\begin{tabular}{ccccccc}
\hline \hline
Revs. &  Date  &  Expo.& SCW&Flux  &Sig.\\
 &  MJD  &  ks & &mCrab &$\sigma$ \\
\hline
0056-0244 & 52729-53291 & 262 & 146  &1.5$\pm$0.3 &4.9\\
\hline
0258-0371 & 53332-53672 & 218  & 129  &1.4$\pm$0.4&3.5\\
\hline
0416-0498 &53805-54050 & 879 & 275 &1.5$\pm$0.2&7.8\\
\hline
0543-0664 & 54184-54547 & 17& 16 &$<$3.3&0.0\\
\hline
0056-0664 & 52729-54547 & 1400& 566 &1.40$\pm$0.15&9.4\\
\hline
\hline
\end{tabular}
\end{center}
\begin{list}{}{}
\item[Note:]{The flux and the significance are presented in the energy band 20-60 keV for the data  in which  the source was within an offset angle of  10 degrees.  The error bars are 1 $\sigma$ and the upper limit is 2 $\sigma$.}
\end{list}
\end{table}

\begin{table}[ptbptbptb]
\begin{center}
\label{tab1}
\caption{ \emph{Swift}/XRT observations log of IGR J19405-3016. }
\vspace{5pt}
\small
\begin{tabular}{cccccc}
\hline \hline
Model & \multicolumn{4}{c} {Observation (date, exposure,ID)}  \\
Par &  2007-07-31  &  2007-08-07 & 2008-06-07&2007-2008 \\
 &  7.6 ks  &  5.9 ks & 1.6 ks& 15.1 ks\\
 &  0036657004  &  0036657005 & 0036657006&all \\
\hline
$\Gamma$&2.07$\pm$0.05 & 2.11$\pm$0.04 & 2.19$^{+0.14}_{-0.13}$ & 2.11$\pm$0.03\\
N & 3.98$\pm$0.13 &5.95$\pm$0.17 & 3.16$\pm$0.26&4.70$\pm$0.10\\
flux & 1.69$^{+0.05}_{-0.06}$ & 2.44$\pm$0.07& 1.19$\pm$0.10&1.94$\pm$0.04\\
$\chi^2$/dof & 1.11/95& 1.12/104&1.16/15 &1.13/198\\
\hline
\hline
\end{tabular}
\end{center}
\begin{list}{}{}
\item[Note:]{ The parameters  $\Gamma$ (photon index) and N (Normalization at 1 keV, in units of 10$^{-3}$ ph cm$^{-2}$ s$^{-1}$ keV$^{-1}$) are the fit results at 0.3-7 keV, using a power law model. The column density is fixed at 8.73$\times$10$^{20}$ atoms/cm$^2$. The $\chi^2$/dof is shown as well for each fit. The flux is calculated  in the 0.3--10 keV band, in units of 10$^{-11}$ erg cm$^{-2}$ s$^{-1}$.}
\end{list}
\end{table}

\begin{figure}[ptbptbptb]
\centering
 \includegraphics[angle=0, scale=0.4]{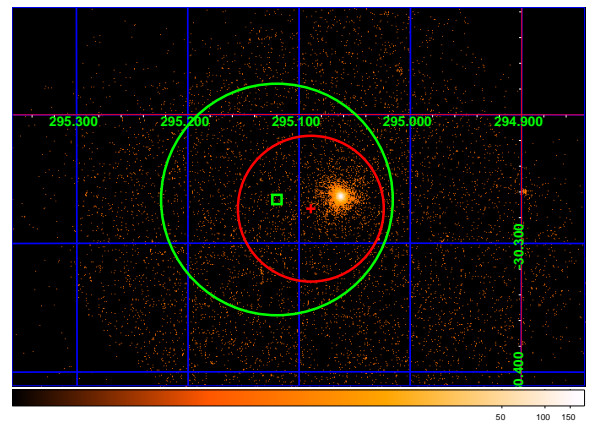}
      \caption{\emph{Swift}/XRT sky map of IGR J19405-3016 detected in  the three combined XRT observations. The hot spot shows the soft X-ray source detected by \emph{Swift}/XRT and the color bar shows  the  scale associated with the image counts.  The open square (green) and the cross symbols (red) are the
locations of IGR J19405-3016 estimated in Bird et al. (2007) and in the current work, respectively.   The two circles are the  corresponding 90\% confidence level error locations. } 
         \label{xrt-map}
\end{figure}

\begin{figure}[ptbptbptb]
\centering
 \includegraphics[angle=0, scale=0.4]{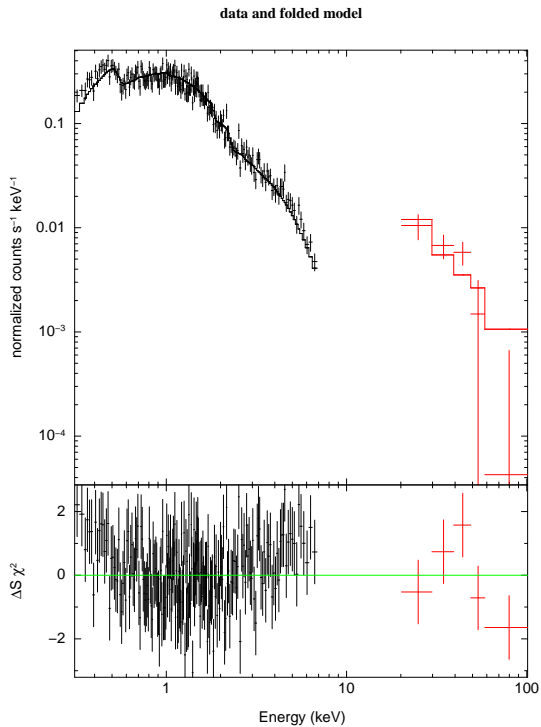}
      \caption{ Upper panel: the combined XRT/ISGRI energy spectrum of IGR J19405-3016 obtained from the  \emph{INTEGRAL} observations of the revolutions 0056 - 0664 and from  the three combined  \emph{Swift}/XRT observations (ID:  00036657004, 00036657005, 00036657006).  The data are  fitted by using a power law shape plus an absorption component fixed at 8.73$\times$10$^{20}$ atoms/cm$^2$.  The residuals (lower panel) show the quality of the fit.  }
         \label{spectrum}
\end{figure}

\acknowledgements
 This work was subsidized by the National Natural Science Foundation of China,  the CAS key Project KJCX2-YW-T03, and the 973 Program 2009CB824800. J.-M. W. thanks the Natural Science Foundation of China for support via NSFC-10325313, 10521001, and 10733010. DFT acknowledges support by Spanish MEC grant AYA 2
006-00530.

\end{document}